\documentclass[superscriptaddress,twocolumn,aps,prl,floatsfix]{revtex4}

\usepackage{ucs}
\usepackage[utf8]{inputenc}
\usepackage[english,francais]{babel}
\usepackage{color}
\usepackage{graphicx}
\usepackage{amssymb,amsmath}
\usepackage{array}
\usepackage{dcolumn}
 \usepackage[normalem]{ulem}

\allowdisplaybreaks

\begin{document}

\title{Analysis of the multiferroicity in the hexagonal manganite
    $\rm YMnO_3$}

\author{Kiran Singh} 
\affiliation{CRISMAT, CNRS-ENSICAEN, Caen, France}
\author{Marie-Bernadette Lepetit$^*$} 
\affiliation{Institut Néel, CNRS, Grenoble, France}
\affiliation{Institut Laue Langevin, Grenoble, France}
\altaffiliation{Previously at CRISMAT, CNRS-ENSICAEN, Caen, France}
\author{Charles Simon$^*$}
\affiliation{Institut Laue Langevin, Grenoble, France}
\altaffiliation{Previously at CRISMAT, CNRS-ENSICAEN, Caen, France}
\author{Natalia Bellido}
\affiliation{CRISMAT, CNRS-ENSICAEN, Caen, France}
\author{Stéphane Pailhès}
\affiliation{Laboratoire Léon Brillouin
CNRS-CEA, CEN Saclay,  Gif/Yvette, France}
\affiliation{IML  University Claude Bernard Lyon I, Villeurbanne, France}
\author{Julien Varignon}
\affiliation{CRISMAT, CNRS-ENSICAEN, Caen, France}
\author{Albin De Muer} \affiliation{LNCMI, CNRS-Université de Grenoble
  Joseph Fourier, Grenoble, France}


\begin{abstract}
  We performed magnetic and ferroelectric measurements, associated
  with Landau theory and symmetry analysis, in order to clarify the
  situation of the $\rm YMnO_3$ system, a classical example of type I
  multiferroics. We found that the only magnetic group compatible with
  all experimental data (neutrons scattering, magnetization,
  polarization, dielectric constant, second harmonic generation) is
  the $P6'_3$ group. In this group a small ferromagnetic component
  along {\bf c} is induced by the Dzyaloshinskii-Moriya interaction,
  and observed here in SQUID magnetization measurements.  We found
  that the ferromagnetic and antiferromagnetic components can only be
  switched simultaneously, while the magnetic orders are functions of
  the polarization square and therefore insensitive to its sign.
\end{abstract}

\maketitle
\section{Introduction}
Hexagonal $\rm YMnO_3$ presents ferroelectricity and
antiferromagnetism~\cite{Bertaut_struct,Smolens_FE_AFM} and can be
considered as the prototype of “type I” ferroelectric
antiferromagnetic materials in which the details of the
magnetoelectric coupling can be studied.  

Despite numerous investigations since the pioneer work of Yakel {\em
  et al.}  in 1963~\cite{Bertaut_struct}, the exact crystalline and
magnetic structures are still under debate. The temperature of the
ferroelectric (FE) transition is for example not completely
clear. Located by some authors at 920K~\cite{Tc=920K}, recent X rays
measurements proposed 1258K~\cite{Gibbs_FE}. These discrepancies are
not fully understood and are possibly due to some changes in the
oxygen deficiency when the sample is heated. Despite these
discrepancies, we can try to summarize the knowledge of this
ferroelectric transition as follows. (i) A transition corresponding to
a unit-cell tripling and a change in space group from centrosymmetric
$P6_3/m mc$ (\#194) to polar $P6_3cm$ (\#185) is observed in this
temperature range. In this respect $\rm YMnO_3$ is a typical example
of an improper ferroelectric~\cite{improp,Moise}, opening the field
to the new concept of hybrid improper ferroelectricity~\cite{hybridFE}. (ii)
Indeed, the symmetric group $P6_3/mmc$ reduces to $P6_3cm$ by a rotation of
the $\rm MnO_5$ polyhedra. A displacement of the yttrium atoms with
respect to the manganese atoms along the c axis of the structure
induces a {\bf c} axis polarization~\cite{Katsu_neutrons,VAken_FE}.
(iii) Furthermore, a possible intermediate phase with the space group
$P6_3/mcm$ can be derived from group theory~\cite{Abraham_FE}, however
it was not observed in the recent measurements~\cite{Gibbs_FE},
neither confirmed by symmetry-mode analysis~\cite{Moise}. The authors
rather observe some evidence for an iso-symmetric phase transition at
about 920~K, which involves a sharp decrease in the estimated
polarization. This transition correlates with several previous reports
of anomalies in physical properties in this temperature
region~\cite{Choi}, but is not really understood.

\begin{figure}[h]
  \begin{center}
    {\Large \bf $V_1$}
    \resizebox{6cm}{!}{\includegraphics{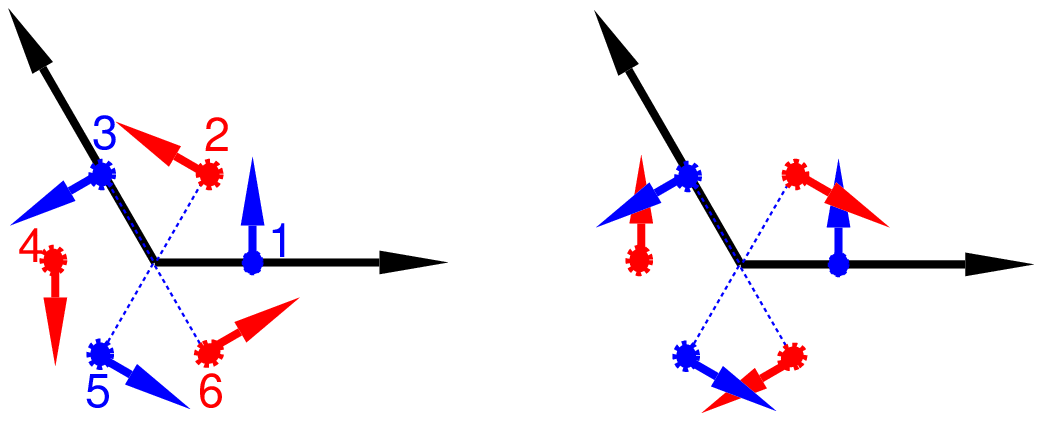}} 
    {\Large \bf $W_1$} \\
    {\Large \bf $V_2$}
    \resizebox{6cm}{!}{\includegraphics{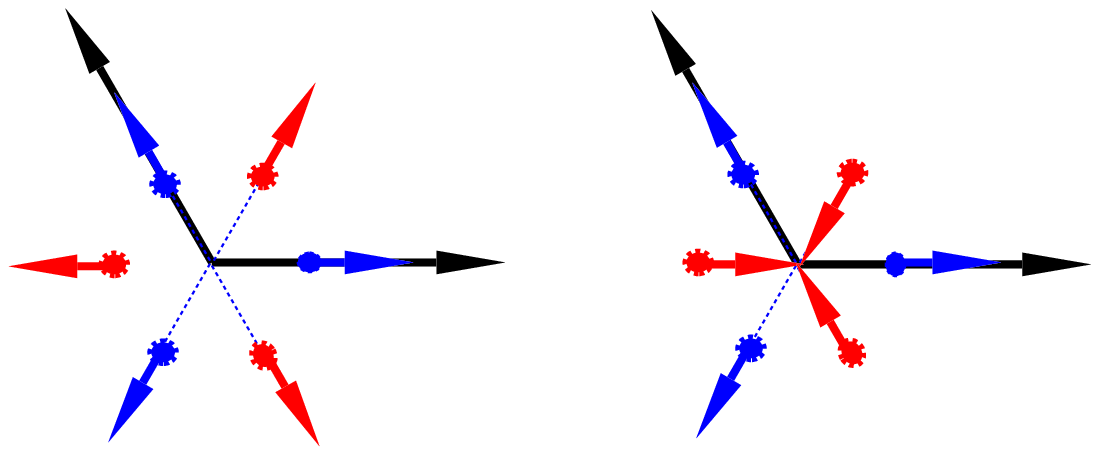}} 
    {\Large \bf $W_2$} \\
    {\Large \bf $V_3$}
    \resizebox{6cm}{!}{\includegraphics{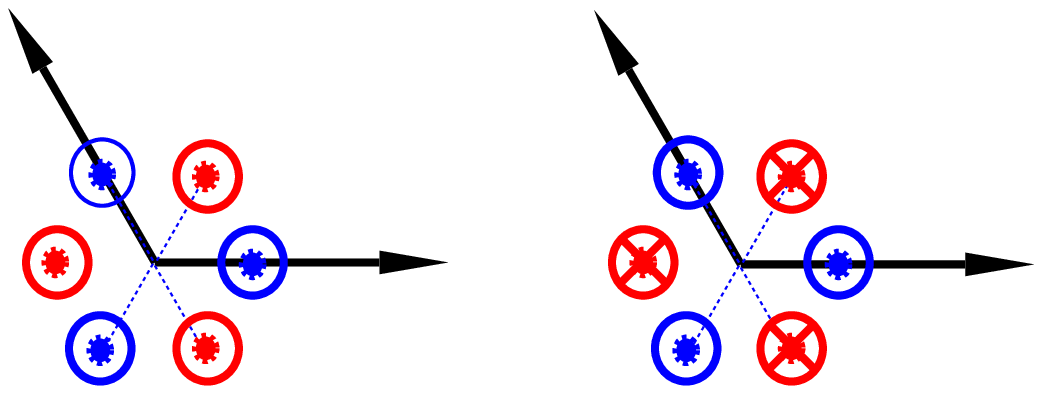}} 
    {\Large \bf $W_3$} 
  \end{center}
  \caption{Schematic representation of the different magnetic orders
    within a unit cell. The Mn sites numbering within the unit cell
    are as follow $\rm Mn_1:(x_{\rm Mn},0,0)$, $\rm Mn_2:(x_{\rm Mn},x_{\rm Mn},1/2)$, $\rm
    Mn_3:(0,x_{\rm Mn},0)$, $\rm Mn_4:(-x_{\rm Mn},0,1/2)$, $\rm
    Mn_5:(-x_{\rm Mn},-x_{\rm Mn},0)$, $\rm Mn_6:(0,-x_{\rm Mn},1/2)$, where $x_{\rm Mn}\simeq 1/3$.}
  \label{fig:VW}
\end{figure}

At $T_N=74\rm\,K$, $\rm YMnO_3$ undergoes a paramagnetic (PM) to
antiferromagnetic (AFM) transition. The magnetism arises from $\rm
Mn^{3+}$ ions, in $3d^4$ configuration, with spins equal to 2 (high
spin). Neutrons diffraction
measurements~\cite{Bertaut_magn1,Lee,Munoz_neutrons,Tapan} showed that
the structure is antiferromagnetic with moments in the {\bf
  ab}-plane. Following Bertaut {\em et al.}~\cite{Bertaut_magn2},
Muñoz {\em et al.}~\cite{Munoz_neutrons}, proposed for the symmetry of
the antiferromagnetic order the $\Gamma_1$ (totally symmetric)
irreducible representation of the $P6_3cm$ group~; this order
corresponds to the$V_1$ order pictured in figure~\ref{fig:VW}. More
recently, a spin polarized analysis showed that the group is rather
$P6_3$ or $P6_3'$~\cite{Tapan}. Finally, in a second harmonic optical
generation work, Fröhlich {\em et al.}  rather concluded to a very
different order associated with the $P6_3’cm’$ magnetic
group~\cite{Frohlich_SHG,Fiebig_SHG}~; this order corresponds to the
$W_2$ order pictured in figure~\ref{fig:VW}. Let us note that, while
Bertaut {\em et al.} and Mu\~noz {\em et al.}  performed a full
symmetry analysis, checking all possible irreducible representations
for the magnetic ordering, Brown and Chatterji, as well as Fiebig {\em
  et al.} only considered the $\Gamma_1$ representation of the tested
symmetry groups. One should however remember that the
magnetic order is the spin part of the system wave-function and as
such can belong to any of the irreducible representation of the
magnetic symmetry group. On another hand, the
polarization behaves as the density matrix and thus can only belong to
the totally symmetric $\Gamma_1$ representation in groups with only
one-dimensional irreducible representations.

Associated with the AFM order, several authors reported a
ferromagnetic (FM) component associated with a spin canting along the
{\bf c} direction. First suggested~\cite{Bertaut_magn1}, and observed
by Bertaut {\em et al.}~\cite{Bertaut_magn2}, this FM component was
later observed in the isotypic compound $\rm ScMnO_3$ by Xu {\em et
  al.}~\cite{Xu_FM_Sc} as well as Bieringer and
Greedan~\cite{Bieringer_FM_Sc}. Attributed to $\rm Mn_3O_4$ impurities
by Fiebig {\em et al}~\cite{Fiebig_SHG}, a FM component disappearing at
$T_N$ was later observed in neutrons scattering by one of
us~\cite{Pailhes_FM}. The controversy about the existence of such a
component is thus still opened. One could argue that the weakness of
the proposed canting removes most of the interest of its existence,
however as we will see in the present paper the existence of a FM
component has many consequences on the symmetry group of the magnetic
structure as well as the interpretation of the $\rm YMnO_3$
properties.

Let us finally quote the existence of a giant magneto-elastic coupling
observed by powder neutron diffraction at the magnetic
transition~\cite{Park08,Patnaik}. Very large atomic displacements (up to
0.1\AA\ ) are induced by the magnetic ordering without any identified
change of the symmetry group. The influence of such displacements on
the polarization or dielectric constant in the magnetic phase was
however never reported on single crystal (such measurements exist in
thin films) while this information is crucial for the assertion of the
assumed magneto-electric coupling seen by domain imaging using second
harmonic generation measurements~\cite{Fiebig_domaines}. 

The present paper aims at building a coherent description for the
magnetic structure of the $\rm YMnO_3$ compound, which will account
for all the experimental observations and resolve their apparent
contradictions.

\section{Can we get some further insight from 
the experiments?}

\subsection{Experimental details}
All the measurements reported in the present work were performed on
the same single crystal, grown long time ago in Groningen by
G. Nénert, from the group of T. Palstra. The sample size for
dielectric measurements is $\rm a=1.1\,mm$, $\rm b=1.5\,mm$ and $\rm
c=0.3\,mm$. Magnetic measurements were performed with a QD MPMS-5
SQUID magnetometer. Dielectric and polarization measurements were
respectively performed in a QD PPMS-14 with Agilent 4284A LCR meter and Keithley
6517A. Magnetic fields above 14\,T (and up to 25\,T) were
achieved in the LNCMI Grenoble. The experimental setup for the
dielectric constant measurements was the same as in Caen, while the
LNCMI setup was used for the magnetization. Antiferromagnetic neutron
diffraction peaks were measured on 4F triple axis spectrometer in
Laboratoire Léon Brillouin in Saclay on the same single crystal.

\subsection{The antiferromagnetic transition}
\begin{figure}[h]
  \centerline{
  \resizebox{8cm}{!}{\includegraphics{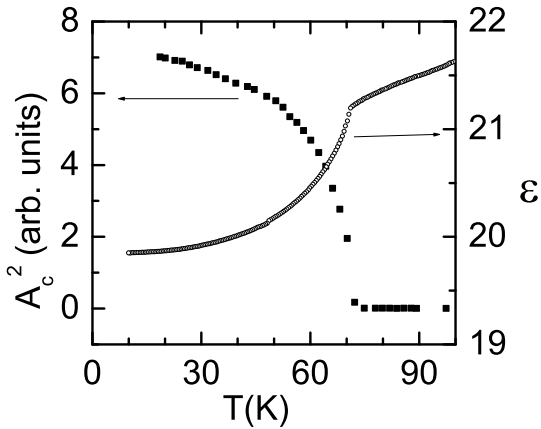}}}
\caption{Temperature dependence of the 100 antiferromagnetic peak
  intensity (left scale) and the ab component of the dielectric
  constant (right scale).}
  \label{fig:neutrons}
\end{figure}

We performed neutron scattering experiments on a neutron triple axis
spectrometer and checked the crystal orientation and crystalline
quality. The 100 magnetic peak is associated with the
antiferromagnetic order parameter. On fig.~\ref{fig:neutrons}, the
temperature dependence of its amplitude is reported, showing the
magnetic transition at $T_N=74\rm\,K$. On the same figure, we reported
the {\bf ab} component of the dielectric constant, $\varepsilon$,
which presents an anomaly at $T_N$. Let us note that the {\bf c}
component of $\varepsilon$ does not present any anomaly at this
temperature (not shown).  The strong similarity, below $T_N$, between
the temperature dependence of the antiferromagnetic order parameter,
and the non linear part of $\varepsilon$, suggests that they are
closely related, and thus infers the existence of a
magneto-electric coupling.  One should emphasize
the fact that the anomaly of the dielectric constant is not a
divergence as  expected in the case of a linear
magneto-electric coupling. This proof of a non-linear magneto-electric
coupling  is of utter importance as
we will see in the next section.

\subsection{The polarization and the dielectric constant}
This magneto-electric coupling can also be asserted from the
polarization and dielectric constant measurements in the magnetic
phase. 

\begin{figure}[h]
 \centerline{
  \resizebox{8cm}{!}{\includegraphics{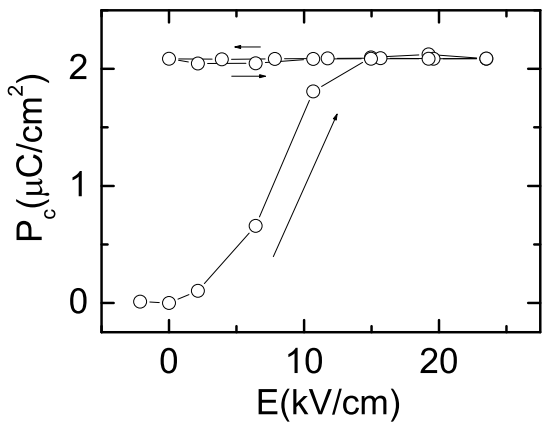}}}
  \caption{Polarization cycle at 30\,K after cooling in a zero
    electric field (the polarization is very small after this
    procedure). The second branch of the measurement ensures that the
    measured current is not due to leakage (the polarization is
    already switched so no change is observed as the electric field is
    switched on again).}
  \label{fig:pol}
\end{figure}

We performed polarization measurements along the {\bf c}-axis (the
only one allowed by symmetry).  A strong reduction of the polarization
amplitude is observed below $T_N$ (see fig.~\ref{fig:pol})~: $2 \rm \,
\mu C/cm^2$ at 30\,K, to be compared with the $5.5 \rm \,\mu C/cm^2$
measured at room temperature~\cite{Kim_pol}.
These polarization values are compatible with the estimated ones,
obtained both as $\vec P=\sum_iq_i\vec r_i$ and from our first
principle calculations.  We computed the polarization using density
functional theory and a Berry phases approach at the atomic structures
given in reference~\onlinecite{Park08} at 10K and 300K.  The
calculations were performed with the B1PW hybrid functionals that was
specifically designed for the treatment of ferroelectric
oxides~\cite{B1PW}. At 300\,K we found a polarization of $5\rm\,\mu
C/cm^2$. in full agreement with experimental values.  At 10\,K, the
polarization is strongly reduced to $1\rm\,\mu C/cm^2$ to be compared
with the experimental result of $2 \rm \, \mu C/cm^2$ at 30\,K.

\begin{figure}[h]
  \centerline{
    \resizebox{8cm}{!}{\includegraphics{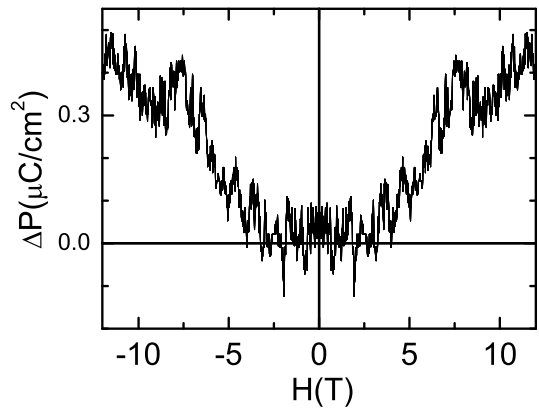}}}
  \caption{Magnetic field dependence of the polarization at 71\,K. }
  \label{fig:polH}
\end{figure}

In addition, we measured the polarization versus the magnetic
field. Since this effect is expected to be very small, we used a
procedure  consisting in ramping many times the magnetic
field from $-14\rm\,T$ to $+14\rm\,T$ and extracting the periodic
signal from the raw data.  One can see on $P(H)$ taken in the magnetic
phase (fig.~\ref{fig:polH}) an anomaly that can be associated with a
meta-magnetic transition.

\begin{figure}[h]
  {\LARGE (a)} \hspace*{6.5cm} \\[-2eM]
  \resizebox{8cm}{!}{\includegraphics{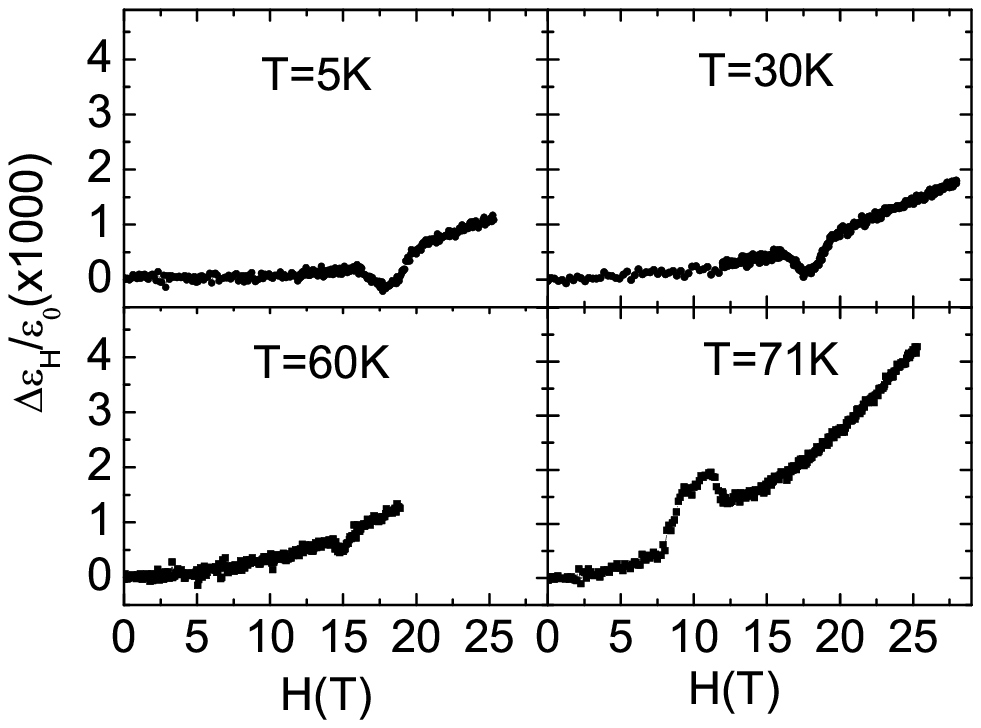}}\\
  {\LARGE (b)} \hspace*{6.5cm} \\[-2eM]
  \resizebox{8cm}{!}{\includegraphics{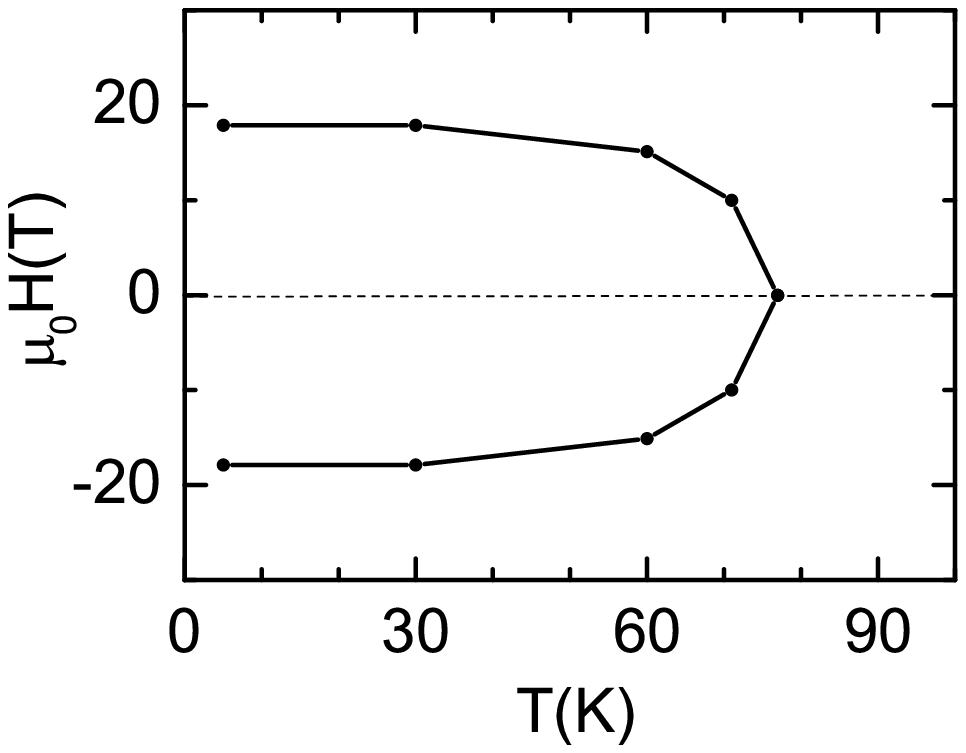}}
  \caption{(a) Magnetic field dependence of the {\bf ab} dielectric
    constant (H is along the {\bf c} axis). (b) Corresponding phase
    diagram (H,T). }
  \label{fig:eps_H}
\end{figure}

This anomaly can also be followed on the dielectric constant,
$\varepsilon$, as a function of applied field and temperature.  The
meta-magnetic transition phase diagram, characteristic of an
antiferromagnetic compound under magnetic field, can
  so be built (see fig.~\ref{fig:eps_H}).
The searched of the meta-magnetic transition is a classical method to
observe the antiferromagnetism. Indeed, in usual systems, the
magnetization (or magnetic susceptibility) versus magnetic field
presents an anomaly at the AFM/FM transition under applied field. In
$\rm YMnO_3$ our ability to see this transition on electric degrees of
freedom (polarization and dielectric constant) clearly proves the
existence of a coupling between the polarization and the magnetic order
parameter, as first proposed by Huang {\em et
  al.}~\cite{Huang}. 

\subsection{The ferromagnetic component}
As mentioned in the introduction, one of us (S. Pailh\`es) observed in
a non-polarized neutrons scattering experiment, a Bragg peak that was
associated with a ferromagnetic component~\cite{Pailhes_FM}. Indeed,
this Bragg peak, at $\vec q=(2,-1,1)$, can neither be associated with
the antiferromagnetic order within the ({\bf a},{\bf b}) plane, nor
with the nuclear extinction rules, since for the $P6_3cm$ symmetry
group imposes $2l=0$. In addition it disappears at $T_N$, as expected
from a canted AFM order. One objection can however be made against
this interpretation. The existence of two $\rm MnO_3$ layers per unit
cell (respectively at $z=0$ and $z=1/2$) , forbids to rule out the
possibility of an antiferromagnetic coupling between the {\bf c}
components of the canted magnetic moments of each layer ($W_3$ versus
$V_3$ order of figure~\ref{fig:VW}).

\begin{figure}[h]
  \centerline{\resizebox{8cm}{!}{\includegraphics{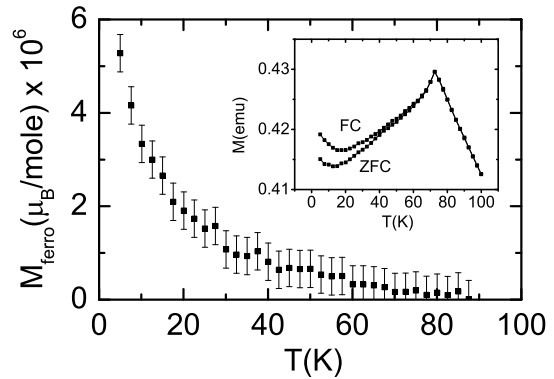}}}
  \caption{Ferromagnetic component {\bf M} as a function of
    temperature. In the inset, the raw data in field cooled (FC) and
    zero field cooled (ZFC) modes. The sample mass is 31.5\,mg.}
  \label{fig:FM}
\end{figure}

We thus performed precise magnetic measurements on a SQUID
magnetometer at low magnetic field, and we did observe a small FM
component (see fig.~\ref{fig:FM}).  The sample was cooled down from
100\,K (still above $T_N$) to 10\,K either under an applied magnetic
field along the {\bf c} axis of the crystal (Field Cooled =FC) or
without any field (Zero Field Cooled = ZFC). After cooling, the
magnetization was always measured ramping the temperature up under the 
applied field. This procedure, assuming that the applied field is too
small to reverse the magnetization, clearly evidenced the existence of
a ferromagnetic component along the {\bf c} axis (see
fig.~\ref{fig:FM}).   The applied magnetic
field is 0.05\,T.

\section{Can we build a theoretical description compatible 
with the  above experimental facts?}
Let us summarize the facts we learned from experiments.
\begin{itemize} \itemsep 1ex
\item $\rm YMnO_3$ exhibits a magneto-electric coupling between the
  AFM and the FE orders.

\item This magneto-electric coupling is non linear. The immediate
  consequence of this is that the AFM order parameter cannot be in the
  same irreducible representation than the FE order parameter, that is
  the polarization. The latter being in the totally symmetric
  irreducible representation~: $\Gamma_1$,  the AFM order cannot belong to
  the $\Gamma_1$ irreducible representation of the magnetic
  group. Assuming that the magnetic order found by
  Bertaut~\cite{Bertaut_magn1,Bertaut_magn2} and
  Mu\~noz~\cite{Munoz_neutrons} is correct, it means that the magnetic
  group is not $P6_3cm$ as assumed by these authors. See $V_1$
  of fig.~\ref{fig:VW} for a picture of this order.

\item There is a weak FM component along the {\bf c} axis.

\item Even if essentially quenched by the crystal field splitting of
  the Mn $3d$ orbitals, the spin-orbit coupling and thus the
  Dzyaloshinskii–Moriya (DM) interaction always exists provided it is
  symmetry allowed. This is the case with the AFM magnetic order $V_1$
  found in neutrons scattering, since the spins vorticity is non
  nil. The DM interaction should thus induces a FM component (even if
  small) along the {\bf c} direction.

\item Finally the AFM and FM order parameters are not linearly
  coupled. Indeed, they present different behaviors around the
  transition (see figure~7 of ref.~\onlinecite{Pailhes_FM} or
  figures~\ref{fig:neutrons} and \ref{fig:FM} of the present paper).
\end{itemize}

According to the above analysis the $P6_3cm$ group cannot be the
system magnetic group. Can we find a magnetic subgroup of the
crystallographic group $P6_3cm$ compatible with all the above
experimental requirements? The following symmetry group analysis tells
us that only one magnetic group is compatible with (i) the $V_1$ AFM
order, (ii) the fact that this order is not in the $\Gamma_1$
irreducible representation and (iii) the existence of a FM component
along the {\bf c} direction. This group is the $P6_3'$ magnetic group.
Indeed, we first examined the magnetic groups associated with the
$P6_3cm$ crystallographic group, that is
\begin{description}
\item[$P6_3cm$~:] discarded since the AFM order $V_1$ belongs to
  $\Gamma_1$ and the FM component $V_3$ is not allowed (does not
  belong to the same representation as $V_1$);
\item[$P6_3'c'm$~:] discarded since $V_1$ and $V_3$ do not
  belong to the same representation (FM component not allowed);
\item[$P6_3'cm'$~:] discarded since the FM component is not allowed;
\item[$P6_3c'm'$~:] discarded since the FM component is not allowed.
\end{description}
Since none of them is compatible with the experimental requirements we
looked further in their subgroups and thus abandoned the mirror planes.
\begin{description}
\item[The $\bf P6_3$ group]  was discarded since $V_1$ belongs to  
  $\Gamma_1$, which is incompatible with the absence of a linear
  magneto-electric coupling.
\item[Finally the $\bf P6_3'$ group] is the only group compatible with
  all the requirements.
\end{description}
 Let us remember that the $P6_3'$ magnetic group was
strongly suggested by Brown and Chatterji~\cite{Tapan} from the
polarimetric study of neutron diffraction. In fact, they were the
first to suggest that the mirror planes are incompatible with the $\rm
YMnO_3$ magnetic group.

Let us now see whether we can account for all the experimental results in a
Landau analysis. We established that the magnetic transition should be
a transition between the paramagnetic (PM) phase belonging to the
$P6_3cm$ group, and the antiferromagnetic (AFM) phase belonging to
the $P6_3'$ group. In the $P6_3'$ group the $\Gamma_4$ irreducible
representation, to whom both the AFM ($V_1$) and the FM ($V_3$) order
parameters belong, is three times represented, namely by the $V_1$,
$V_2$ and $V_3$. The Landau theory must thus involve all three
magnetic order parameters in addition to the change in the
ferroelectric polarization. $V_1$ and $V_2$ are easily represented by
the toroidal ($\vec A$) and divergence ($B$) components of the
in-plane spins component, while $V_3$ is the out of plane component
associated with the magnetization ($\vec M$). For each unit cell one
can thus define
\begin{eqnarray*} 
\vec A &=& \frac{1}{6r} \sum_i \vec r_i \wedge \vec S_i = 
\frac{1}{6r} \sum_i \vec r_i \wedge \vec S_{ab,i} \\
B &=&  \frac{1}{6r} \sum_i  \vec r_i \cdot \vec S_i =  
\frac{1}{6r} \sum_i  \vec r_i \cdot \vec S_{ab,i} \\
\vec M &=&  \frac{1}{6} \sum_i   \vec S_i = \frac{1}{6} \sum_i   \vec S_{c,i} 
\end{eqnarray*}
where the summations over $i$ run over the six Mn atoms of the unit
cell~; the $\vec r_i$ refer to the in plane components of the Mn atoms
position vectors (note that $\sum_i \vec r_i = \vec 0$ and $\forall i\,
|\vec r_i|=r$)~; the $\vec S_i$ are the Mn atomic spins ($\vec S_i =
\vec S_{ab,i} + \vec S_{c,i}$ where $\vec S_{ab,i}$ is the in-plane
component of the Mn spins and $\vec S_{c,i}$ is the {\bf c} axis
component). 

$\vec A$ and $\vec M$ are vectors along the {\bf c} direction while
$B$ is a scalar. Let use write $A = S_{ab}\cos{\varphi}$ and
$B=S_{ab}\sin{\varphi}$ and point out that the intensity of the 100
AFM magnetic peak (fig.~\ref{fig:neutrons}) is proportional to
${S_{ab}}^2$ whatever the angle $\varphi$.  In the paramagnetic state,
i.e. for $T > T_N$, $S_{ab} = 0$, but the polarization $P$ is not
zero. This is one of the important issue of this compound. $\vec P$ is
not a driving order parameter for the magnetic transition~; however,
since its value presents a singularity at $T_N$, it is a secondary
order parameter. Its contribution should thus be taken into account in
the Landau free energy and can only contain even powers of $P$, as
imposed by the higher temperature paraelectric to ferroelectric
transition. The free energy can thus be expressed up to the power 4 of
the order parameters 
\begin{eqnarray*}
  F &=& \overbrace{\alpha_2(T-T_N) (A^2+B^2) + \alpha_4(A^2+B^2)^2}^{\text{AFM energy}} \\ 
  && \overbrace{-\beta_2(P^2-{P_0}^2) + \beta_4(P^4-{P_0}^4)}^{\text{change in the FE energy}} \\ 
  && \overbrace{ + \gamma_2 M^2 + \gamma_4 M^4}^{\text{FM energy}} \\ 
  && \overbrace{ + c_4 (A^2+B^2)(P^2-{P_0}^2)}^{\text{AFM/FE coupling}} \\ 
  && \overbrace{ + d_4 M^2(P^2-{P_0}^2)}^{\text{FM/FE coupling}}  \\ 
  && \overbrace{ + e_4 (A^2+B^2)M^2}^{\text{AFM/FM coupling}} \\
  && \overbrace{ + z_4 (P^2-{P_0}^2) \vec A \cdot \vec M}^{\text{Dzyaloshinskii–Moriya interaction}} 
\end{eqnarray*}
where $\alpha_2$, $\alpha_4$, $\beta_2$, $\beta_4$, $\gamma_2$,
 $c_4$, $d_4$, $e_4$, $z_4$ are the temperature independent Landau
expansion coefficients. If one notes $t=T_N-T$, and $\vec{\delta
  P}=\vec P-\vec P_0$ the gradient of the free energy writes as
\begin{eqnarray*}
  \frac{\partial F}{\partial S_{ab}} &=&
  S_{ab}\left[   -2\alpha_2\,t + 4\alpha_4\,{S_{ab}}^2 
    + 2c_4  \delta\!P\,(2P_0+\delta\!P) \right.  \\ && \left.
    + 2e_4 M^2 \right]
  + z_4 \, \cos{\varphi}\,M\,   \delta\!P\,(2P_0+\delta\!P) =0 \\[2ex] 
  \frac{\partial F}{\partial\delta\!P} &=&
  (P_0+\delta\!P) \left[ -2\beta_2
    + 4\beta_4\,({P_0}^2 + 2{P_0}\delta\!P  +{\delta\!P}^2) 
    \right. \\ && \left.
    + 2c_4 \, {S_{ab}}^2  
    + 2d_4 \, M^2  \quad
    + 2z_4 \,  S_{ab}\,M\,\cos{\varphi}\right] = 0 \\[2ex]  
  \frac{\partial F}{\partial \varphi} &=&
  z_4 \,  S_{ab}\,M\,\sin{\varphi}\, \delta\!P\,(2P_0+\delta\!P) = 0 \\[2ex] 
  \frac{\partial F}{\partial M} &=&
  M\,\left[ 2\gamma_2  + 4\gamma_4 M^2 
    + 2d_4 \,\delta\!P\,(2P_0+\delta\!P) 
    + 2e_4 {S_{ab}}^2 \right] \\ && + 
  z_4 \, S_{ab} \,  \cos{\varphi}\, \delta\!P\,(2P_0+\delta\!P) = 0
\end{eqnarray*}

From the experimental results we know that $M \ll {S_{ab}}^2$ and
$\delta\!P$. We thus expect that if $S_{ab}\propto t^\nu$,
$\delta\!P\propto t^\mu$ and $M\propto t^\eta$, we will have in the
vicinity of the transition $\eta > \mu$ and $\eta > 2\nu$. In an order
by order expansion of the free energy gradient as a function of $t$,
one can thus suppose either that $\eta > \nu+1$ ($M\ll S_{ab}t$) or
that $\eta \sim \nu+1$ ($M\sim S_{ab}t$). It is easy to show that the
first hypothesis leads to a contradiction. Let us thus assume that
$M\sim S_{ab}t$. One gets at the zeroth order in $t$
\begin{eqnarray*}
  \frac{\partial F}{\partial\delta\!P} =
  P_0 \left[ -2\beta_2  + 4\beta_4\,{P_0}^2  \right]  = 0 
&\quad \Leftrightarrow \quad& P_0^2 = \frac{\beta_2}{2\beta_4}
\end{eqnarray*}
and at the following order
\begin{eqnarray*}
&& \left\{
    \begin{array}[c]{lcl}
      \frac{\partial F}{\partial S_{ab}}  &: \quad &
      S_{ab}\left[   -2\alpha_2\,t + 4\alpha_4\,{S_{ab}}^2 \
        + 4c_4P_0  \delta\!P \right] =0 \\[1ex]
      \frac{\partial F}{\partial\delta\!P}  &: \quad &
      8\beta_4\,{P_0}^2\delta\!P   + 2c_4 P_0\, {S_{ab}}^2  = 0
    \end{array} \right.  \\[2ex]
 & \Leftrightarrow &
 \left\{
    \begin{array}[c]{lcl}
      {S_{ab}}^2 &=& \frac{\alpha_2\,\beta_2}{2\alpha_4\,\beta_2   -  {c_4}^2{P_0}^2}\,t \\[1ex]
      \delta\!P &=& - \frac{\alpha_2\,c_4P_0}{4\alpha_4\,\beta_2   -  2{c_4}^2{P_0}^2}\,t  
    \end{array} \right.  \\
\end{eqnarray*}
\begin{eqnarray*}
  \frac{\partial F}{\partial \varphi}  &: \quad &
  2z_4 P_0\, S_{ab}\, M\,\sin{\varphi}\, \delta\!P = 0 
  \quad \Leftrightarrow \quad
  \sin{\varphi}=0 \\[2ex] 
  \frac{\partial F}{\partial M}  &: \quad &
  2\gamma_2  M +   2z_4P_0 \, S_{ab} \,  \cos{\varphi}\, \delta\!P = 0
  \\ &\Leftrightarrow &
  M = 
  \frac{{\alpha_2}^{3/2}\,{\beta_2}^{1/2} c_4z_4{P_0}^2}
  {2\gamma_2\,(2\alpha_4\,\beta_2   -  {c_4}^2{P_0}^2)^{3/2}}
     \,\cos{\varphi}\,t^{3/2}  
\end{eqnarray*}
We thus retrieve the $V_1$ order for the AFM spins arrangement
($\sin{\varphi}=0$) ~; the decrease in the polarization amplitude
under the N\'eel transition ($\delta P<0$), the fact that the FM order
parameter is much weaker than both the AFM one and the change in the
polarization ($\nu=1/2$, $\mu=1$, $\eta=3/2$), and
finally the fact the FM and AFM order parameters are not linearly
related at $T_N$.  The polarization and the square of the AFM order
parameter are predicted to vary linearly in $t$ at the magnetic
transition, as a classical second order phase transition. In fact, as
it is for most magnetic phase transitions, higher order terms in the
free energy make the temperature dependence over a large scale of
temperature different from the mean field prediction. Here for
example, the best fit for ${S_{ab}}^2$ is a power law in $t^{1/3}$
(not shown in fig.~\ref{fig:neutrons}).

Coming back to the anomaly of the dielectric constant at $T_N$ and using
the second derivative of $F$ with respect to $P$, one gets in the
first order in $t$
\begin{eqnarray*}
  \frac{1}{\chi_e} &=& \frac{\partial^2 F}{\partial{\delta\!P}^2} =
    4\beta_2 - 6 \, c_4\, S^2   + 2c_4 \, S^2  
   = 4 (\beta_2 - c_4\, S^2)  
   \\[2ex]
\text{and}\quad   \varepsilon &=& 1+\chi_e\\
   &=& 1 + \frac{1}{4\beta_2} + \frac{c_4}{4{\beta_2}^2}\, S^2 
\end{eqnarray*}
Comparing the above expression with the experimental data of
fig.~\ref{fig:neutrons}, the Landau analysis correctly predicts the
critical shape of $\varepsilon$ versus the AFM order
parameter ${S_{ab}}^2$.

As a first conclusion one can state that the above Landau analysis
seems in perfect agreement with all the experimental data.  The most
important consequence of it is that one cannot switch the direction of
any of the magnetic orders ---~clockwise vs counter clockwise rotation
of the antiferromagnetic order (sign of $A$) or direction of the
magnetization (sign of $M$)~--- by switching $P$. Indeed, one has $P =
P_0 \left(1- \frac{c_4}{4\beta_4} \, A^2\right)$, and $ M =
-\frac{z_4}{\gamma_2}\, A \, P_0\delta\!P$ thus a change in the sign
of $P$ will  leave the sign of both $A$ and $M$ unchanged. On the
contrary, $A$ and $M$ are switched simultaneously.

\section{Are there other options?}
If one supposes that the weak FM component is artefactual, then there
are three different groups compatible with the $V_1$ AFM order and the
absence of a linear magneto-electric coupling, that is~: $P6_3'c'm$,
$P6_3'cm'$ and $P6_3c'm'$. In such a case however it is difficult to
explain why the $\vec D_{ij}.(\vec S_i\wedge \vec S_j)$
Dzyaloshinskii-Moriya interaction does not yield a FM component along
the $\bf c$ direction. Giving up the FM component thus means giving up
the $V_1$ ordering for the AFM order. 

Is there another AFM order compatible with the neutrons scattering
experiments? Following Bertaut~\cite{Bertaut_magn3} and
Mu\~noz~\cite{Munoz_neutrons}, there is indeed
another AFM order possibly compatible with the neutrons diffraction
data, even if with a significantly worse agreement factor than $V_1$
($R_\text{MAG}=10.8\%$ instead of $7.6\%$~\cite{Munoz_neutrons}). This
order is  pictured as $W_2$ in
fig.~\ref{fig:VW}.  It is compatible with a non-linear
magneto-electric coupling in the $P6_3cm$, $P6_3'c'm$, $P6_3c'm'$ and
$P6_3$ magnetic groups. In the $P6_3cm$, $P6_3'c'm$ and $P6_3c'm'$
groups it is associated in its irreducible representation with the
$W_3$ order, while in the $P6_3$ magnetic groups both the $W_1$ and
$W_3$ orders belong to the representation of $W_2$. At this point let
us note that the $W_3$ AFM order is compatible with the $(2,-1,1)$
peak observed by Pailh\`es {\it et al}~\cite{Pailhes_FM} in neutrons
scattering.

\subsection{What Landau's theory tells us~? }
In the $P6_3cm$, $P6_3'c'm$ and $P6_3c'm'$ groups the Landau analysis yields 
\begin{eqnarray*}
  F &=& \alpha_2(T-T_N) B'^2 + \alpha_4B'^4 \quad + \quad
  \gamma_2 M'^2 + \gamma_4 M'^4\\ 
  && -\beta_2(P^2-{P_0}^2) + \beta_4(P^4-{P_0}^4) \\ 
  &&  + c_4 B'^2(P^2-{P_0}^2)
      + d_4 M'^2(P^2-{P_0}^2)
      + e_4 B'^2M'^2
\end{eqnarray*}
where $B'$ and $M'$ are the order parameters respectively associated
with $W_2$ and $W_3$.
\begin{eqnarray*}
  B' &=& \frac{1}{6r} \sum_i (-1)^i \vec r_i \cdot \vec S_i =
\frac{1}{6r} \sum_i (-1)^i \vec r_i \cdot \vec S_i^{a,b} \\
\vec M' &=&  \frac{1}{6} \sum_i (-1)^i \vec S_i = 
\frac{1}{6} \sum_i (-1)^i \vec S_{c,i} 
\end{eqnarray*}
\begin{eqnarray*}
  \frac{\partial F}{\partial S_{ab}} &=&
  S_{ab}\left[   -2\alpha_2\,t + 4\alpha_4\,{S_{ab}}^2 
    + 2c_4  \delta\!P\,(2P_0+\delta\!P) \right.  \\ && \left.
    + 2e_4 M'^2 \right]
   =0 \\[2ex] 
  \frac{\partial F}{\partial\delta\!P} &=&
  (P_0+\delta\!P) \left[ -2\beta_2
    + 4\beta_4\,({P_0}^2 + 2{P_0}\delta\!P  +{\delta\!P}^2) 
    \right. \\ && \left.
    + 2c_4 \, {S_{ab}}^2  
    + 2d_4 \, M'^2     \right] = 0 \\[2ex]  
  \frac{\partial F}{\partial M'} &=&
  M'\,\left[ 2\gamma_2  + 4\gamma_4 M'^2 
    + 2d_4 \,\delta\!P\,(2P_0+\delta\!P) \right. \\ && \left.
    + 2e_4 {S_{ab}}^2 \right] 
  = 0
\end{eqnarray*}
These equations give
\begin{eqnarray*}
     P_0^2 &=& \frac{\beta_2}{2\beta_4} \\ 
     {S_{ab}}^2 = \frac{\alpha_2\,\beta_2}{2\alpha_4\,\beta_2 - {c_4}^2{P_0}^2}\,t 
     &\text{and}&
     \delta\!P = -\frac{\alpha_2\,c_4P_0}{4\alpha_4\,\beta_2-2{c_4}^2{P_0}^2}\,t 
     \\[1ex] 
     \text{and finally} \qquad M'&=& 0
\end{eqnarray*}

One sees that these results are equivalent to the previous derivation
as far as $S_{ab}$ and $P$ are concerned. However, if $M'=0$ is
coherent with the neutrons scattering results  of
ref.~\cite{Munoz_neutrons}, it is not compatible with the existence of
the $(2,-1,1)$ peak observed by Pailh\`es {\it et
  al}~\cite{Pailhes_FM}. Indeed, in this representation the $(2,-1,1)$
peak measures the intensity of the order parameter $M'$.

\subsection{What about the second harmonic generation experiments?}
\begin{figure}[h]
    \centerline{\resizebox{6cm}{!}{\includegraphics{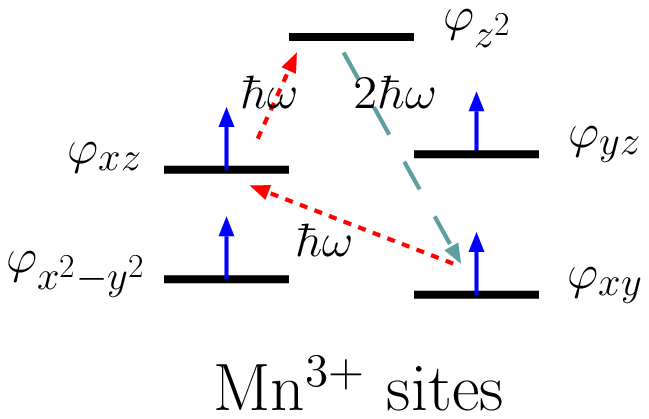}}}
  \caption{Example of the two photons absorption (in dotted red) and
    one photon emission (in dashed blue-grey) responsible for the second
    harmonic generation.}
  \label{fig:shg}
\end{figure}

The second harmonic spectra are due to $d$--$d$ electronic transitions
within the $\rm Mn^{3+}$ ions (see figure~\ref{fig:shg}). The non
linear succeptibility is dominated by the term starting from the
atomic ground state ($S=2$) and can be written as
\begin{eqnarray*}
 \epsilon_0 \; \chi_{\alpha\beta\gamma} &=& 
  \sum_{m,k}\frac{ \langle 0|\hat P_\alpha|m\rangle 
    \langle m|\hat P_\beta|k\rangle 
    \langle k|\hat P_\gamma|0\rangle }
  {( \hbar\omega_{mi}-2 \hbar\omega)( \hbar\omega_{ki}- \hbar\omega)}    
\end{eqnarray*}
where $|0\rangle$ is the $\rm Mn^{3+}$ atomic ground state~;
$|k\rangle$ and $|m\rangle$ span the $d$--$d$ $\rm Mn^{3+}$ excited
states, $\hbar\omega_{ki}$ and $\hbar\omega_{mi}$ being their
excitation energies~; $\hat P_\alpha$ are the dipolar moment operators
along the $\alpha$ direction.

We thus evaluated both the in-plane, $\epsilon_0 \; \chi_{xxx}$ and
$\epsilon_0 \; \chi_{yyy}$, and the out-of-plane components,
$\epsilon_0 \; \chi_{zxx},\,\epsilon_0 \; \chi_{zyy}$, for the
different magnetic groups and orders discussed in this paper. The
detailed calculations can be found in Appendix.

For the magnetic groups associated with the $P6_3cm$ space group we found 
{\small  \begin{eqnarray} \label{eq:chi_P63cm}
    \epsilon_0 \, \chi_{xxx}(z\!=\!0) &=&
    \epsilon_0 \, \chi_{xxx}^0(z\!=\!0) 
    \\ && \hspace{-15ex}
    + \left[ 
      \frac{A}{\varepsilon_{z^2}-\varepsilon_{x^2-y^2}-2\hbar\omega}
      + \frac{A'}{\varepsilon_{z^2}-\varepsilon_{xy}-2\hbar\omega}\right]
    \langle S_x\rangle_{{\rm Mn}_1}
    \nonumber \\ && \hspace{-15ex}
    + \left[ 
      \frac{B}{\varepsilon_{z^2}-\varepsilon_{x^2-y^2}-2\hbar\omega}
      + \frac{B'}{\varepsilon_{z^2}-\varepsilon_{xy}-2\hbar\omega}\right]
    \langle S_x\rangle_{{\rm Mn}_1}^2 
    \nonumber \\[2ex]
    \epsilon_0 \, \chi_{yyy}(z\!=\!0) &=& 0
    \nonumber \\[2ex]
    \epsilon_0 \, \chi_{zxx}(z\!=\!0) &=&   
    \epsilon_0 \, \chi_{zxx}^0(z\!=\!0) 
    + \frac{C}{\varepsilon_{z^2}-\varepsilon_{x^2-y^2}-2\hbar\omega}
    \langle S_x\rangle_{{\rm Mn}_1}
    \nonumber \\ && \hspace{-18ex}
    + \frac{D}{\varepsilon_{z^2}-\varepsilon_{x^2-y^2}-2\hbar\omega}
    \langle S_x\rangle_{{\rm Mn}_1}^2 
    + \frac{E}{\varepsilon_{z^2}-\varepsilon_{x^2-y^2}-2\hbar\omega}
    \langle S_x\rangle_{{\rm Mn}_1}^3 \nonumber
  \end{eqnarray}
} and similarly for the $z\!=\!1/2$ layer ($\rm Mn_1$ being replaced
by $\rm Mn_4$).  $\chi_{xxx}^0(z\!=\!0)$ is the spin independent (FE)
tensor, $\varepsilon_{i}$ the energy of the iron orbital
$\varphi_i\simeq 3d_i$ and $\langle S_\alpha\rangle_{{\rm Mn}_i}$ the
$\alpha$ component of the spin of the $\rm Mn_i$ atom.  $x,y,z$ are
orthogonal axes, $x$ being along the crystallographic $a$ direction
and $k$ along the crystallographic $c$ direction.

From these results one can derive the following conclusions.
\begin{itemize}
\item Within the symmetry rules associated with a $P6_3cm$ crystal
  group the second harmonic signal can only be sensitive to magnetic
  orders in which $\langle S_x\rangle_{{\rm Mn}_1}\ne 0$ and/or
  $\langle S_x\rangle_{{\rm Mn}_4}\ne 0$.
\item The experimental data~\cite{Frohlich_SHG} that sees a magnetic
  contribution to the in-plane component of $\chi$ are thus
  incompatible with the $V_1$ magnetic order as previously shown by
  Iizuka-Sakano~\cite{Ii_SHG} and coherently with our previous analysis.
\item We showed that the only possible magnetic order compatible with
  a $P6_3cm$ crystal group is $W_2$ in which $\langle
  S_x\rangle_{{\rm Mn}_1} = \langle S_x\rangle_{{\rm Mn}_4}\ne
  0$. According to equations~\ref{eq:chi_P63cm} this order predicts a
  magnetic contribution to the in-plane component of $\chi$, but also
  to the out-of-plane one $\chi_{zxx}=\chi_{zyy}$. While the first one
  is in agreement with the experimental findings, no magnetic signal
  was found in the out-of-plane SHG signal. 
\item The $P6_3cm$ crystal group and associated magnetic groups are
  thus not only incompatible with the existence of a FM component and
  the $(2,-1,1)$ peak observed by neutrons scattering but also with
  the SHG experimental data.
\end{itemize}

Let us thus go back to the $P6_3'$ magnetic group and remember that,
up to now, this group was found compatible with all experimental
facts. The calculation yields the following form for the $\chi$
components (only the contributions associated with the $V_1$ and $V_3$
magnetic orders compatible with the $P6_3'$ magnetic group are
retained) {\small
\begin{eqnarray*}
  \epsilon_0  \, \chi_{xxx}(z=0)&=& \epsilon_0 \, \chi_{xxx}^0(z=0) 
  \\ && \hspace{-18ex} 
  +  {\bf i }\left[
    \frac{A}{\varepsilon_{z^2}-\varepsilon_{x^2-y^2}-2\hbar\omega} 
    + \frac{A'}{\varepsilon_{z^2}-\varepsilon_{xy}-2\hbar\omega} 
  \right]
  \langle S_y\rangle_{\rm Mn_1} \langle S_z\rangle_{\rm Mn_1}
  \\ && \hspace{-18ex} 
  +  \left[
    \frac{B}{\varepsilon_{z^2}-\varepsilon_{x^2-y^2}-2\hbar\omega}
    + \frac{B'}{\varepsilon_{z^2}-\varepsilon_{xy}-2\hbar\omega}
  \right] \langle S_y\rangle_{\rm Mn_1}^2
  \\[2ex]
  \epsilon_0  \, \chi_{yyy}(z=0)&=& \epsilon_0 \, \chi_{yyy}^0(z=0) 
  \\ && \hspace{-16ex} 
  +  \left[
    \frac{C}{\varepsilon_{z^2}-\varepsilon_{x^2-y^2}-2\hbar\omega} 
    + \frac{C'}{\varepsilon_{z^2}-\varepsilon_{xy}-2\hbar\omega} 
  \right]
   \langle S_y\rangle_{\rm Mn_1}
  \\ && \hspace{-17ex} 
  + {\bf i}\, \left[   \left(
      \frac{D}{\varepsilon_{z^2}-\varepsilon_{x^2-y^2}-2\hbar\omega}
      + \frac{D'}{\varepsilon_{z^2}-\varepsilon_{xy}-2\hbar\omega}
    \right)  \langle S_y\rangle_{\rm Mn_1}^2
  \right. \\ && \left. \hspace{-14.6ex} + 
    \left(
      \frac{E}{\varepsilon_{z^2}-\varepsilon_{x^2-y^2}-2\hbar\omega}
      + \frac{E'}{\varepsilon_{z^2}-\varepsilon_{xy}-2\hbar\omega}
    \right) 
  \right]
  \langle S_z\rangle_{\rm Mn_1}
  \\[2ex]
  \epsilon_0  \chi_{zxx}(z=0)&=& \epsilon_0  \chi_{zyy}(z=0)
  \;=\; 
  \epsilon_0 \chi_{zxx}^0 
  \\ && \hspace{-14ex} 
  \;+\;   \left[
   \frac{A}{\varepsilon_{z^2}-\varepsilon_{x^2-y^2}-2\hbar\omega}
   + \frac{A'}{\varepsilon_{z^2}-\varepsilon_{xy}-2\hbar\omega}
   \right] \langle S_y\rangle_{\rm Mn_1}^2
\end{eqnarray*}
} Let us remember that the $V_1$ and $V_3$ orders cannot be reversed
independently ($\langle S_y\rangle_{\rm Mn_1}\langle S_z\rangle_{\rm
  Mn_1} >0$ whatever the magnetic domain), and that $\langle
S_y\rangle_{\rm Mn_1}=-\langle S_y\rangle_{\rm Mn_4}$, $\langle
S_z\rangle_{\rm Mn_1}=\langle S_z\rangle_{\rm Mn_4}$. One thus sees
immediately that $\chi_{xxx}$ and $\chi_{zxx}$ depend only on $\langle
S_y\rangle_{\rm Mn_1}^2$ and should thus be insensitive to the
magnetic domains. On the contrary, $\chi_{yyy}$ depend on $\langle
S_z\rangle_{\rm Mn_1}$ and should thus exhibit a sensitivity to the
magnetic domains at two different frequencies~; namely
$2\hbar\omega=\varepsilon_{z^2}-\varepsilon_{xy}$ and
$2\hbar\omega=\varepsilon_{z^2}-\varepsilon_{x^2-y^2}$, differing by
$\Delta E'$. Those results are in full agreement with the experimental
data reported on reference~\onlinecite{Frohlich_SHG}.

\section{Conclusion}
In the present paper we showed from joined experimental evidences and
theoretical analysis that the AFM transition in $\rm YMnO_3$ is
associated with three order parameters, namely the AFM one (primary
order parameter), the extra-component of the polarization along {\bf
  c} and the ferromagnetic component along the {\bf c} axis induced by
the Dzyaloshinskii-Moriya interaction (secondary order
parameters). Moreover the analysis of the magnetic transition showed
the absence of linear coupling between them and thus a hierarchy.
Taking into account the different experimental observations (magnetic
and transport macroscopic measurements, neutrons scattering data,
optical second harmonic responses), as well as the presence of the DM
coupling, it appears that the $P6_3'$ magnetic group is the only
possible one. In the past, many publications tried to address this
question with different conclusions, but all of them present unsolved
questions or problems we tried to address in the present work.  For
example, the importance of a ferromagnetic component was underlined by
Bertaut, but corresponds in his samples to a parasitic phase~; some
authors have discarded magnetic groups, assuming that the magnetic
order should belong to the $\Gamma_1$ irreducible representation of
the symmetry group, and so forgetting that despite being by far the
most frequent, this is not the only possibility and any of the group
representation is valid for the wave function. In fact, the absence of
a divergence in the dielectric constant at the phase transition
implies that the magneto-electric coupling is not linear, and thus
that the polarization and the AFM order cannot belong to the same
irreducible representation. The polarization being of $\Gamma_1$
symmetry, the magnetic order cannot belong to the totally symmetric
representation $\Gamma_1$. This, in addition to the presence of the
small ferromagnetic component, implies that the only possible group is
here $P6_3'$. In this group, a change in the sign of the polarization,
P, will let both the weak magnetization, M, and the AFM order
parameter, A, unchanged. On the contrary, A and M will be switched
simultaneously. For possible applications, this type of multiferroic
cannot be used to switch the magnetization with an electric field, but
rather to switch antiferromagnetism with an intense magnetic field,
providing memories which are only little sensitive to magnetic fields.

\section*{Acknowledgments}
The authors thank G. N\'enert and TM. Palstra for providing them with the
sample, the IDRIS and CRIHAN French computer centers for providing
them with computer time.

\section*{Appendix}
\subsection*{General considerations}
In the following appendix the SHG equations are expressed in term of
an orthogonal $x,y,z$ set of axes. The $x$ axis is along the $a$
direction, that is associated with one of the O--Mn bonds in the $z=0$
layer (O is the $(0,0,z\simeq0)$ in-plane oxygen), the $y$
axis its in-plane orthogonal and the $z$ axis is along the $c$ direction.

The three-fold rotation axis is present in any
of the groups proposed in this paper. We can thus use it in order to
express the $\chi_{\alpha\beta\gamma}$ tensor for the $z=0$ layers as
a function of its value for the $\rm Mn_1$ ion (see fig.~\ref{fig:VW}
for the ions labeling), and for the $z=1/2$ layer as a function of its
value for the $\rm Mn_4$ ion. One gets easily 
\begin{eqnarray*}
  \chi_{xxx}(z=0) &=&
  \frac{3}{4}\left(\chi_{xxx}({\rm Mn}_1)- \sum\chi_{xyy}({\rm Mn}_1)\right)\\
  \chi_{zxx}(z=0) &=& 
  \frac{3}{2}\left(\chi_{xxz}({\rm Mn}_1)+ \chi_{yyz}({\rm Mn}_1)\right)\\
  \chi_{xzz}(z=0) &=& 0 
\end{eqnarray*}
and similarly for $z=1/2$ with $\rm Mn_4$ or for $\chi_{yyy},
\chi_{zyy}, \chi_{yzz}$.  The summation $\sum\chi_{xyy}$ must be
intended as a sum over all similar terms, that is $\sum\chi_{xyy}=
\chi_{xyy} + \chi_{yxy} + \chi_{yyx}$. 

Starting from the high temperature phase, we will proceed in
perturbation (up to the first order in the wave functions, second
order in energy) to include the different symmetry breaking at the FE
and AFM transitions, as well as the spin-orbit interaction.  In the
$P6_3/mmc$ high temperature group, the Mn ions are located on sites of
$D_{3h}$ symmetry and one gets the following $3d$ zeroth order
orbitals (associated with a nil non linear succeptibility tensor)
\begin{eqnarray*}
  \varphi_{z^2} &=& d_{z^2}   \\
  \varphi_{xz} &=& d_{xz}    \\
  \varphi_{yz} &=& d_{yz}   \\
  \varphi_{x^2-y^2} &=&  c d_{x^2-y^2} + c' p_x  \\
  \varphi_{xy} &=&  c d_{xy} + c'p_y
\end{eqnarray*}
At this point let us notice that the $d_{x^2-y^2}$ and $p_x$ (as well
as the $d_{xy}$ and $p_y$) Mn orbitals belong to the same irreducible
representation and are thus hybridized through the metal-ligand
interactions.

\subsection*{The magnetic groups associated with the $P6_3cm$ 
crystal group}
Going through the FE transition toward the $P6_3cm$ group, the Mn ions
goes from a $D_{3h}$ site to a $C_s$ symmetry site, thus the
degeneracies between $\varphi_{xz}\,/\,\varphi_{yz}$ and the
$\varphi_{x^2-y^2}\,/\,\varphi_{xy}$ orbitals are lifted by
respectively $\delta E$ and $\delta E'$.  At the first order of
perturbation in this symmetry breaking and in the spin orbit
coupling, one gets the following orbitals
\begin{eqnarray} \label{eq:orbcm}
  \varphi_{z^2} & =& d_{z^2} + \varsigma p_z 
  + \mu d_{xz} + \nu (c d_{x^2-y^2} +  c' p_x) 
    \nonumber \\ && 
  + \frac{\sqrt{3}\aleph}{\Delta\varepsilon_2}
  \left[\langle S_x\rangle d_{xz} + \langle S_y\rangle d_{yz}\right]
  \nonumber \\[2ex]
  \varphi_{xz} & =& d_{xz} + \lambda (c d_{x^2-y^2} + c' p_x) 
  - \mu d_{z^2} +  \varsigma' p_z \nonumber \\ && 
  -  \frac{\sqrt{3}\aleph}{\Delta\varepsilon_2} \langle S_x\rangle d_{z^2}
  + i\frac{\aleph}{\delta E}\langle S_z\rangle d_{yz}
  \nonumber \\ && 
  + \frac{c\aleph}{\Delta\varepsilon_1}
  \left[\langle S_x\rangle (cd_{x^2-y^2}+c' p_x) 
    + \langle S_y\rangle (cd_{xy}+ c'p_y)\right]\nonumber \\[2ex]
  \varphi_{yz} & =& d_{yz} + \lambda (c d_{xy} +  c' p_y) 
  \nonumber \\ && 
  - \frac{\sqrt{3}\aleph}{\Delta\varepsilon_2}\langle S_y\rangle d_{z^2}
  + i\frac{\aleph}{\delta E}\langle S_z\rangle d_{xz}
  \nonumber \\ && 
  + \frac{c\aleph}{\Delta\varepsilon_1}
  \left[ -\langle S_y\rangle (cd_{x^2-y^2}+c' p_x) 
    + \langle S_x\rangle (cd_{xy}+ c'p_y)\right] \nonumber \\[2ex]
  \varphi_{x^2-y^2} &=&  c d_{x^2-y^2} + c' p_x - \lambda  d_{xz} - \nu d_{z^2}
  +  \varsigma'' p_z  
 \nonumber \\ && 
  - \frac{c \aleph}{\Delta\varepsilon_1} 
  \left[\langle S_x\rangle d_{xz} - \langle S_y\rangle d_{yz}\right]
  \nonumber \\ && 
  + i  \frac{c^2 2 \aleph}{\delta E'}
  \langle S_z\rangle (cd_{xy}+ c'p_y) \nonumber \\[2ex]
  \varphi_{xy} &=&  c d_{xy} + c'p_y - \lambda d_{yz}\nonumber \\ && 
  -  \frac{c\aleph}{\Delta\varepsilon_1}
  \left[\langle S_y\rangle d_{xz} + \langle S_x\rangle d_{yz}\right]
  \nonumber \\ && 
  + i\frac{ c^2 2 \aleph}{\delta E'}\langle S_z\rangle (cd_{x^2-y^2}+c' p_x)
\end{eqnarray}
where $\aleph$ is the spin-orbit coupling constant, and $\langle S_j
\rangle$ the average values of the spin operators associated with
ground state spin order.
$\Delta\varepsilon_1$ is the excitation energy from the degenerated
$\varphi_{xz}$ or $\varphi_{yz}$ orbitals toward the $\varphi_{z^2}$
one, $\Delta\varepsilon_2$ is the excitation energy from the
degenerated $\varphi_{x^2-y^2}$, $\varphi_{xy}$ orbitals toward the
$\varphi_{xz} $ or $\varphi_{yz}$ ones. $\lambda, \mu, \nu, \varsigma,
\varsigma', \varsigma''$ are the first order mixing coefficients
associated with the $P6_3/mmc \rightarrow P6_3cm$ symmetry breaking.
 
For any of the magnetic groups associated with the $P6_3cm$ spatial
group, the non linear succeptibility tensor will involve the following
transitions (authorized light polarization is shown on top of the
arrows while the orbitals irreps are given in parentheses)
\begin{widetext}
$$ \left\{ \begin{array}[c]{cl}
    \varphi_{xz} (A) \overset{xz}{\longrightarrow} \varphi_{z^2} (A) & 
    \left\{\begin{array}[c]{c@{\quad}l}
       \varphi_{x^2-y^2} (A) \overset{x,z}{\longrightarrow} \varphi_{xz} (A)&
       \varphi_{z^2} (A) \overset{x,z}{\longrightarrow} \varphi_{x^2-y^2} (A)\\ 
       \varphi_{xy} (A') \overset{y}{\longrightarrow} \varphi_{xz} (A)&
       \varphi_{z^2} (A) \overset{y}{\longrightarrow} \varphi_{xy} (A')\\  
     \end{array} \right. \\[3ex]
    \varphi_{yz} (A') \overset{y}{\longrightarrow} \varphi_{z^2} (A) & 
    \left\{\begin{array}[c]{c@{\quad}l}
       \varphi_{x^2-y^2} (A) \overset{y}{\longrightarrow} \varphi_{yz} (A')&
       \varphi_{z^2} (A) \overset{x,z}{\longrightarrow} \varphi_{x^2-y^2} (A)\\ 
       \varphi_{xy} (A') \overset{x,z}{\longrightarrow} \varphi_{yz} (A')&
       \varphi_{z^2} (A) \overset{y}{\longrightarrow} \varphi_{xy} (A')\\  
     \end{array} \right.
 \end{array} \right\} 
$$ 
\end{widetext}

Using the above diagram and the orbitals given in
equations~\ref{eq:orbcm} one can show that 
{\small \begin{eqnarray*} \label{eq:chi_P63cm} \epsilon_0 \,
    \chi_{xxx}(z\!=\!0) &=& \epsilon_0 \, \chi_{xxx}^0(z\!=\!0) \\ &&
    \hspace{-15ex} + \left[
      \frac{A}{\varepsilon_{z^2}-\varepsilon_{x^2-y^2}-2\hbar\omega} +
      \frac{A'}{\varepsilon_{z^2}-\varepsilon_{xy}-2\hbar\omega}\right]
    \langle S_x\rangle_{{\rm Mn}_1} \nonumber \\ && \hspace{-15ex} +
    \left[
      \frac{B}{\varepsilon_{z^2}-\varepsilon_{x^2-y^2}-2\hbar\omega} +
      \frac{B'}{\varepsilon_{z^2}-\varepsilon_{xy}-2\hbar\omega}\right]
    \langle S_x\rangle_{{\rm Mn}_1}^2
    \nonumber \\[2ex]
    \epsilon_0 \, \chi_{yyy}(z\!=\!0) &=& 0
    \nonumber \\[2ex]
    \epsilon_0 \, \chi_{zxx}(z\!=\!0) &=&   
    \epsilon_0 \, \chi_{zxx}^0(z\!=\!0) 
    + \frac{C}{\varepsilon_{z^2}-\varepsilon_{x^2-y^2}-2\hbar\omega}
    \langle S_x\rangle_{{\rm Mn}_1}
    \nonumber \\ && \hspace{-18ex}
    + \frac{D}{\varepsilon_{z^2}-\varepsilon_{x^2-y^2}-2\hbar\omega}
    \langle S_x\rangle_{{\rm Mn}_1}^2 
    + \frac{E}{\varepsilon_{z^2}-\varepsilon_{x^2-y^2}-2\hbar\omega}
    \langle S_x\rangle_{{\rm Mn}_1}^3 \nonumber
  \end{eqnarray*}
}
and similarly for the $z\!=\!1/2$ layer. $\chi_{xxx}^0(z\!=\!0)$ is the spin
independent (FE) tensor and $\varepsilon_{i}$ the energy of orbital $\varphi_i$.

For the $W_2$ magnetic order one has $\langle S_x\rangle_{{\rm
    Mn}_1}=\langle S_x\rangle_{{\rm Mn}_4}$ thus if $\epsilon_0 \,
\chi_{xxx}^0$ and $\chi_{zxx}$ include all the magnetic domain independent terms
{\small \begin{eqnarray*} 
\epsilon_0 \, \chi_{xxx}
    &=& \epsilon_0 \, \chi_{xxx}^0 \\ && \hspace{-15ex} + 2 \left[
      \frac{A}{\varepsilon_{z^2}-\varepsilon_{x^2-y^2}-2\hbar\omega} +
      \frac{A'}{\varepsilon_{z^2}-\varepsilon_{xy}-2\hbar\omega}\right]
    \langle S_x\rangle_{{\rm Mn}_1}
    \nonumber \\[2ex]
    \epsilon_0 \, \chi_{zxx}^0 &=&  
    2\frac{C}{\varepsilon_{z^2}-\varepsilon_{x^2-y^2}-2\hbar\omega}
    \langle S_x\rangle_{{\rm Mn}_1}
    \nonumber \\ && 
    + 2 \frac{E}{\varepsilon_{z^2}-\varepsilon_{x^2-y^2}-2\hbar\omega}
    \langle S_x\rangle_{{\rm Mn}_1}^3 \nonumber
  \end{eqnarray*}
} It results that in this scheme both the in-plane and out-of-plane
signal should be sensitive to the magnetic domains.

On the contrary,  the $V_1$ magnetic order should not
display any SHG signal since $\langle S_x\rangle_{{\rm Mn}_1}=\langle
S_x\rangle_{{\rm Mn}_4}=0$.

\subsection*{The  $P6'_3$ magnetic group}
Let us now look at the $P6_3'$ magnetic group. The associated crystal
group is $P6_3$ in which the Mn ions are on a $P_1$ symmetry site. In
this group the Fe $3d$ orbitals can be expressed as {\small
\begin{eqnarray} \label{eq:orb63} 
  \varphi_{z^2} &=&  d_{z^2} + \varsigma p_z 
  + \mu d_{xz} + \nu (c d_{x^2-y^2} +  c' p_x) 
  \nonumber  \\ && 
  + \mu' d_{yz} 
  + \nu' [c d_{xy} + c'p_y] 
  \nonumber \\ && 
  + \frac{\sqrt{3}\aleph}{\Delta\varepsilon_2}
  \left[\langle S_x\rangle d_{xz} + \langle S_y\rangle d_{yz}\right]
  \nonumber \\[2ex]
  \varphi_{xz} &=&  d_{xz} + \lambda (c d_{x^2-y^2} + c' p_x) 
  - \mu d_{z^2} +  \varsigma' p_z \nonumber \\ && 
  + \tau d_{yz}  + \upsilon [c d_{xy} + c'p_y] \nonumber \\ && 
  -  \frac{\sqrt{3}\aleph}{\Delta\varepsilon_2} \langle S_x\rangle d_{z^2}
  + i\frac{\aleph}{\delta E}\langle S_z\rangle d_{yz}
  \nonumber \\ && 
  + \frac{c\aleph}{\Delta\varepsilon_1}
  \left[\langle S_x\rangle (cd_{x^2-y^2}+c' p_x) 
    + \langle S_y\rangle (cd_{xy}+ c'p_y)\right]
  \nonumber \\[2ex]
  \varphi_{yz} &=& d_{yz} + \lambda (c d_{xy} +  c' p_y)
  \nonumber \\ && 
  -  \mu' d_{z^2}  +  \varsigma''' p_z h  - \tau d_{xz}
  + \upsilon'[ c d_{x^2-y^2} + c' p_x]\nonumber \\ && 
  - \frac{\sqrt{3}\aleph}{\Delta\varepsilon_2}\langle S_y\rangle d_{z^2}
  + i\frac{\aleph}{\delta E}\langle S_z\rangle d_{xz}
  \nonumber \\ && 
  + \frac{c\aleph}{\Delta\varepsilon_1}
  \left[ -\langle S_y\rangle (cd_{x^2-y^2}+c' p_x) 
    + \langle S_x\rangle (cd_{xy}+ c'p_y)\right] \nonumber \\[2ex]
  \varphi_{x^2-y^2} &=&   [c d_{x^2-y^2} + c' p_x] - \lambda  d_{xz} 
  - \nu d_{z^2}  +  \varsigma'' p_z 
  \nonumber \\ && 
  - \upsilon' d_{yz} + \tau'[c d_{xy} + c'p_y] \nonumber \\ && 
  - \frac{c \aleph}{\Delta\varepsilon_1} 
  \left[\langle S_x\rangle d_{xz} - \langle S_y\rangle d_{yz}\right]
  \nonumber \\ && 
  + i  \frac{c^2 2 \aleph}{\delta E'}
  \langle S_z\rangle (cd_{xy}+ c'p_y) \nonumber \\[2ex]
  \varphi_{xy} &=&  [c d_{xy} + c'p_y] - \lambda d_{yz}
  \nonumber \\ && 
  - \nu'd_{z^2} +   \varsigma'''' p_z   - \tau'[c d_{x^2-y^2} + c' p_x]
  - \upsilon d_{xz} \nonumber \\ && 
  -  \frac{c\aleph}{\Delta\varepsilon_1}
  \left[\langle S_y\rangle d_{xz} + \langle S_x\rangle d_{yz}\right]
  \nonumber \\ && 
  + i\frac{ c^2 2 \aleph}{\delta E'}\langle S_z\rangle (cd_{x^2-y^2}+c' p_x)
\end{eqnarray}
}
where $\mu',\nu', \tau, \tau', \upsilon, \upsilon',\varsigma''',
\varsigma''''$ are the first order perturbation coefficients associated
with the $P6_3cm \rightarrow P6_3$ symmetry breaking.  The non linear
succeptibility tensor will thus involve the following transitions
\begin{widetext}
$$\left\{ \begin{array}[c]{clcl}
    \varphi_{xz}  &\overset{x,y,z}{\longrightarrow} &\varphi_{z^2}  & 
    \left\{\begin{array}[c]{clc@{\qquad}clc}
        \varphi_{x^2-y^2} &\overset{x,y,z}{\longrightarrow}& \varphi_{xz} &
        \varphi_{z^2}     &\overset{x,y,z}{\longrightarrow}& \varphi_{x^2-y^2} \\  
        \varphi_{xy}      &\overset{x,y,z}{\longrightarrow}& \varphi_{xz} &
        \varphi_{z^2}     &\overset{x,y,z}{\longrightarrow}& \varphi_{xy}\\  
      \end{array} \right.\\[5ex]
    \varphi_{yz}  &\overset{x,y,z}{\longrightarrow}& \varphi_{z^2}  & 
    \left\{\begin{array}[c]{clc@{\qquad}clc}
        \varphi_{x^2-y^2} &\overset{x,y,z}{\longrightarrow}& \varphi_{yz} &
        \varphi_{z^2} &\overset{x,y,z}{\longrightarrow}& \varphi_{x^2-y^2} \\  
        \varphi_{xy}  &\overset{x,y,z}{\longrightarrow}& \varphi_{yz} &
        \varphi_{z^2} &\overset{x,y,z}{\longrightarrow}& \varphi_{xy} \\  
    \end{array} \right.
\end{array} \right\} 
$$
\end{widetext}
As it is expected that the $P6_3cm \rightarrow P6_3$ punctual symmetry
breaking is very weak (not observed in X-ray diffraction up to now),
in the calculation of the second harmonic succeptibility tensor we
will thus neglect the terms in $\mu',\nu', \tau, \tau', \upsilon,
\upsilon',\varsigma''', \varsigma''''$.  Using the above diagram and
the orbitals given in equations~\ref{eq:orb63} one can show that the
SHG tensor has the following form (only the contributions associated
with the $V_1$ and $V_3$ magnetic orders compatible with the $P6'_3$
magnetic group are retained)
{\small 
\begin{eqnarray*}
  \epsilon_0  \, \chi_{xxx}(z=0)&=& \epsilon_0 \, \chi_{xxx}^0(z=0) 
  \\ && \hspace{-18ex} 
  +  {\bf i }\left[
    \frac{A}{\varepsilon_{z^2}-\varepsilon_{x^2-y^2}-2\hbar\omega} 
    + \frac{A'}{\varepsilon_{z^2}-\varepsilon_{xy}-2\hbar\omega} 
  \right]
  \langle S_y\rangle_{\rm Mn_1} \langle S_z\rangle_{\rm Mn_1}
  \\ && \hspace{-18ex} 
  +  \left[
    \frac{B}{\varepsilon_{z^2}-\varepsilon_{x^2-y^2}-2\hbar\omega}
    + \frac{B'}{\varepsilon_{z^2}-\varepsilon_{xy}-2\hbar\omega}
  \right] \langle S_y\rangle_{\rm Mn_1}^2
  \\[2ex]
  \epsilon_0  \, \chi_{yyy}(z=0)&=& \epsilon_0 \, \chi_{yyy}^0(z=0) 
  \\ && \hspace{-16ex} 
  +  \left[
    \frac{C}{\varepsilon_{z^2}-\varepsilon_{x^2-y^2}-2\hbar\omega} 
    + \frac{C'}{\varepsilon_{z^2}-\varepsilon_{xy}-2\hbar\omega} 
  \right]
   \langle S_y\rangle_{\rm Mn_1}
  \\ && \hspace{-17ex} 
  + {\bf i}\, \left[   \left(
      \frac{D}{\varepsilon_{z^2}-\varepsilon_{x^2-y^2}-2\hbar\omega}
      + \frac{D'}{\varepsilon_{z^2}-\varepsilon_{xy}-2\hbar\omega}
    \right)  \langle S_y\rangle_{\rm Mn_1}^2
  \right. \\ && \left. \hspace{-14.6ex} + 
    \left(
      \frac{E}{\varepsilon_{z^2}-\varepsilon_{x^2-y^2}-2\hbar\omega}
      + \frac{E'}{\varepsilon_{z^2}-\varepsilon_{xy}-2\hbar\omega}
    \right) 
  \right]
  \langle S_z\rangle_{\rm Mn_1}
  \\[2ex]
  \epsilon_0  \chi_{zxx}(z=0)&=& \epsilon_0  \chi_{zyy}(z=0)
  \;=\; 
  \epsilon_0 \chi_{zxx}^0 
  \\ && \hspace{-14ex} 
  \;+\;   \left[
   \frac{A}{\varepsilon_{z^2}-\varepsilon_{x^2-y^2}-2\hbar\omega}
   + \frac{A'}{\varepsilon_{z^2}-\varepsilon_{xy}-2\hbar\omega}
   \right] \langle S_y\rangle_{\rm Mn_1}^2
\end{eqnarray*}
}
Using $\langle S_y\rangle_{\rm Mn_1}\langle S_z\rangle_{\rm
  Mn_1} >0$ whatever the magnetic domain,  $\langle
S_y\rangle_{\rm Mn_1}=-\langle S_y\rangle_{\rm Mn_4}$ and $\langle
S_z\rangle_{\rm Mn_1}=\langle S_z\rangle_{\rm Mn_4}$, one gets
{\small 
\begin{eqnarray*}
  \epsilon_0  \, \chi_{xxx}&=& \epsilon_0 \, \chi_{xxx}^0
  \\[2ex]
  \epsilon_0  \, \chi_{yyy}&=& \epsilon_0 \, \chi_{yyy}^0
  \\ && \hspace{-10ex} 
  + 2{\bf i}\, \left[
      \frac{D  \langle S_y\rangle_{\rm Mn_1}^2+E }
      {\varepsilon_{z^2}-\varepsilon_{x^2-y^2}-2\hbar\omega}
      + \frac{D'\langle S_y\rangle_{\rm Mn_1}^2+E'}
      {\varepsilon_{z^2}-\varepsilon_{xy}-2\hbar\omega}
    \right]  
  \langle S_z\rangle_{\rm Mn_1}
  \\[2ex]
  \epsilon_0  \chi_{zxx}&=& \epsilon_0  \chi_{zyy}
  \;=\; 
  \epsilon_0 \chi_{zxx}^0 
\end{eqnarray*}
}
One sees immediately that $\chi_{xxx}$ and $\chi_{zxx}$ should be insensitive
to the magnetic order, while $\chi_{yyy}$ should exhibit a sensitivity
to the magnetic domains at two different frequencies.


\end{document}